\documentclass[12pt]{article}

\usepackage{arxiv}
\usepackage{setspace}
\usepackage{amsmath}
\usepackage[utf8]{inputenc} 
\usepackage{hyperref}       
\usepackage{url}            
\usepackage{booktabs}       
\usepackage{multicol} 
\usepackage{multirow} 
\usepackage{makecell} 
\usepackage{amssymb}   
\usepackage{amsfonts}
\usepackage{mathrsfs}
\usepackage{placeins} 
\usepackage{bm}  
\usepackage{graphicx}
\usepackage{nicefrac}       
\usepackage{microtype}      
\usepackage{array}      
\usepackage{tcolorbox}
\usepackage[]{algorithm2e}
\usepackage{dutchcal} 
\usepackage[T1]{fontenc}
\usepackage[style=authoryear,natbib]{biblatex}
\newcommand{\norm}[1]{\left\lVert #1 \right\rVert}

\title{How good is good? Probabilistic benchmarks and nanofinance+}

\author{  Rolando Gonzales Martinez\thanks{This research was carried out with funds provided by the FAHU foundation. The author would like to thank the comments of M. A. Mosashvili.} \\ Helmholtz-Zentrum Dresden-Rossendorf (Germany)  | University of Agder (Norway) \\  \texttt{rolando.gonzales@hzdr.de | rolando.gonzales@uia.no} \\}

\begin{document}

\maketitle

\begin{abstract}
 Benchmarks are standards that allow to identify opportunities for improvement among comparable units. This study suggests a 2-step methodology for calculating probabilistic benchmarks in noisy data sets: (\textit{i}) double-hyperbolic undersampling filters the noise of key performance indicators (KPIs), and (\textit{ii}) a relevance vector machine estimates probabilistic benchmarks with denoised KPIs. The usefulness of the methods is illustrated with an application to a database of nano-finance+. The results indicate that---in the case of nano-finance groups---a higher discrimination power is obtained with variables that capture the macro-economic environment of the country where a group operates. Also, the estimates show that groups operating in rural regions have different probabilistic benchmarks, compared to groups in urban and peri-urban areas.
\end{abstract}

\keywords{benchmarks \and nanofinance(+) \and KPIs \and denoising \and relevance vector machines}

\newpage

\section{Introduction}

Benchmarking is the process of analyzing key performance indicators with the aim of creating standards for comparing competing units \citep{bogetoft2010benchmarking}. Probabilistic benchmarks measure the probability of a unit falling into an interval along with the cumulative probability of exceeding a predetermined threshold \citep{wolfe2019diagnosis}.  As a management tool, benchmarks allow to identify and apply better documented practices \citep{bogetoft2013performance}. 

Benchmarks are widely used in diverse scientific disciplines. Pharmaceutics compare the prices of prescription drugs with benchmarks \citep{gencarelli2005one}. In environmental science, benchmarks set water quality standards \citep{van2019specific} or define thresholds for radiation risk \citep{bates2011environmental}. In finance, interest-rate benchmarks mitigate search frictions by lowering informational asymmetries in the markets \citep{duffie2017benchmarks}.


This study develops a 2-step processes for calculating probabilistic benchmarks in noisy datasets. In step 1, double-hyperbolic undersampling filters the noise of key performance indicators (KPIs); in step 2, a relevance vector machine estimate probabilistic benchmarks with filtered KPIs. Archimidean copulas approximate the joint density of KPIs during the denoising step. Besides estimating probabilistic benchmarks, the methods of step 2 identify the continuous and categorical factors influencing benchmarks.


The 2-step methodology is illustrated with an application to a database of nanofinance+ working with business interventions. In nanofinance, low-income individuals without access to formal financial services get together and start to accumulate their savings into a fund, which they later use to provide themselves with loans and insurance. In nanofinance+ (NF+), development agencies, donors and governments help communities to create NF+ groups for financial inclusion and then the groups become a platform for additional `plus' sustainable development programs---see  \citet{gonzales2019b} for details.

The methods proposed in this study complement the state-of-the-art in probabilistic benchmarking of \citet{chakarov2014expectation}, \citet{chiribella2014quantum} or \citet{yang2014certifying}. Along with this methodological contribution, the empirical findings of this document fill the research gap left by economic studies that have been focused only on calculating benchmarks for microfinance institutions---see for example \citet{tucker2001} or \citet{reille2002}. In microfinance, benchmarks are used to compare institutions; in nanofinance, benchmarks are aimed to compare groups. Benchmarks for nanofinance groups allow to set performance standards for monitoring and evaluating intervention programs implemented in communities worldwide.

The definition of multivariate probabilistic benchmarks used in the study is described in Section \ref{sec:def}. Section \ref{sec:methods} discusses the methods for estimating multivariate probabilistic benchmarks in noisy datasets. Section \ref{sec:empap} shows the empirical application to the NF+ database. Section \ref{sec:conclusion} concludes. The data and the MatLab codes that allow to replicate the results of the study are freely available at MathWorks file-exchange  (https://nl.mathworks.com/matlabcentral/fileexchange/74398-double-hyperbolic-undersampling-probabilistic-benchmarks). 


\section{Multivariate probabilistic benchmarks}\label{sec:def}

Classical benchmarking makes use of fixed inputs to calculate point estimates for classification standards. Probabilistic benchmarking, in contrast, takes into account elements of uncertainty in the inputs and thus generates interval estimates as an output \citep{liedtke1998irreproducible}. For example, probabilistic benchmarks are calculated for quantum information protocols---teleportation and approximate cloning---in \citet{yang2014certifying}; more recently, \citet{lipsky2019accuracy} calculate probabilistic benchmarks for noisy anthropometric measures, and \citet{wolfe2019diagnosis} use probabilistic benchmarks to quantify the uncertainty in fibromyalgia diagnosis.

Proposition 1 below shows the definition of multivariate probabilistic benchmarks used in this study. 

\begin{flushleft}\rule{\textwidth}{0.4pt}\end{flushleft}
\noindent\textit{\textbf{Proposition 1: Multivariate probabilistic benchmarks.} Let $\mathbf{y}$ be a $N \times j$ matrix $\mathbf{y} \in \mathbb{R}^j$ of $j$-KPIs ($y_1,y_2,...,y_j$ key performance indicators) for a set $\mathcal{H}$ of $\left\{ \eta_1, \eta_2,..., \eta_N \right\} \ni \mathcal{H}$ comparable units. Given the joint density,
\[
f_{\mathbf{y}} \left(\mathbf{y}\right) := f\left(y_1,y_2,...,y_j\right)  
 = \frac{\partial F\left(y_1,y_2,...,y_j\right)}{\partial y_1 \partial y_2 \cdots \partial y_j},
\]
where $F\left(y_1,y_2,...,y_j\right)$ is a CDF and $f\left(y_1,y_2,...,y_j\right) \geq 0$, the differentiated units $\mathcal{h}_{\tau} \subset \mathcal{H}$ will be those for which:
\begin{equation}
1 - \int_{\tau_1}^{\infty} \int_{\tau_2}^{\infty} \cdots \int_{\tau_j}^{\infty} f\left(y_1,y_2,...,y_j\right) d y_1 d y_2 \cdots d y_j,
\label{eq:succ}
\end{equation}
given a threshold $\tau$ in $\tau \in \mathbb{R}^j$.} \newline
\rule{\textwidth}{0.4pt}\bigskip

In proposition 1, the discrimination of $\eta_1, \eta_2,..., \eta_N$ units in a comparable set $\mathcal{H}$ is based on interval estimates of a multi-dimensional threshold (the benchmark) $\tau$. Proposition 1 sets a probabilistic standard based on the joint multivariate distribution function of the KPIs $\left\{y_1,y_2,...,y_j\right\} \ni \mathbf{y}$ used for calculating $\tau$. The isolines---the contour intervals---defined by the benchmarks $\tau$ allow to identify the units $\mathcal{h}_{\tau}$ with a different performance in the unit hypercube ($\mathcal{h}_{\tau} \subset \mathcal{H}$).


Proposition 2 below states that the thresholds $\tau$ can be calculated without the need to know the exact form of the joint density $f_{\mathbf{y}} \left(\mathbf{y}\right)$ in Equation \ref{eq:succ}:

\begin{flushleft}\rule{\textwidth}{0.4pt}\end{flushleft}
\noindent\textit{\textbf{Proposition 2: Unit-hypercube approximation.}
Let $\mathcal{C}_{\Theta}:[0,1]^d \mapsto [0,1]$ be a $d$-dimensional multivariate cumulative distribution function with $\mathbf{u} \in \left\{u_1, u_2, ... , u_d\right\}$ uniform marginal distributions and a dependence structure defined by $\Theta$. If $\mathbf{u} \equiv F_{\mathbf{y}} \left( \mathbf{y}\right)$,  the joint density of $\mathbf{y}$ needed to calculate $\tau$ can be approximated with the simulation of $\mathcal{C}_{\Theta} \left( \mathbf{u} \right)$ in the unit hypercube:
\[
\mathcal{C}_{\Theta} \left( \mathbf{u} \right)
:=
\mathcal{C}_{\Theta} \left( u_1, u_2,\dots,u_d\right) = \int_{-\infty}^{u_1} \int_{-\infty}^{u_2} \cdots \int_{-\infty}^{u_j} c \left( u_1, u_2, \dots, u_j\right) d u_1 d u_2 \cdots d u_j,
\]
for $\mathcal{C}_{\Theta} \left( \mathbf{u} \right) = 0$ if $u_d = 0$, $\mathcal{C}_{\Theta} \left( \mathbf{u} \right) = u_d$ for any $u_d = 1$, and $\mathcal{C}_{\Theta}\left(\cdot\right)$ satisfies the non-negativity condition on the volume, i.e. $\mathcal{C}_{\Theta}\left(\cdot\right)$ is $d$-increasing quasi-monotone in $[0,1]^j$. 
}\newline
\rule{\textwidth}{0.4pt}\bigskip

Proposition 2 is based on Sklar's theorem (\cite{sklar1959fonctions}; \cite{sklar1996random}), which indicates that any multivariate joint distribution can be written in terms of univariate marginal distribution functions and a copul{\ae} $\mathcal{C}_{\Theta}$ that captures the co-dependence between the variables \citep{DURANTE2013945}.

Archimidean copulas are a type of copul{\ae} that approximate the joint multivariate distribution of KPIs that are not elliptically distributed  \citep{naifar2011modelling}. In an Archimedean copula $\mathcal{C}_g$, an additive generation function $g(u)$ models the strength of dependence in arbitrarily high dimensions with only one scalar parameter: $\theta$ \citep{smith2003modelling}. Formally:
\begin{equation}
   \mathcal{C}_g (u_1, u_2, ... , u_d) =
    \begin{cases}
        g^{-1}(g(u_1) + g(u_1) + \cdots + g(u_d)) \text{ if } \sum_{v= 1}^d  g(u_v) \leq g(0) \\
        0 \text{ otherwise, }
    \end{cases}
\end{equation}
with $g(u)$ a generator function that satisfies $g(1) = 0$, $g'(u) < 0$ and $g''(u) > 0$ for all $0 \leq u \leq 1$; hence $\mathcal{C}_{\theta} \equiv \mathcal{C}_g$. In Clayton's Archimedean copula, for example, the generator function is equal to $g_{\theta} (u) = u^{-\theta} -1$ for $\theta > 1$ (\cite{mcneil}; \cite{cherubini2011dynamic}): 
\begin{equation}
    \mathcal{C}_{\theta} (u_1, u_2, ... , u_d) =
    \left(
    1 - d + \sum_{v= 1}^d u_v ^{-\theta}
    \right)^{-1/\theta}.
\end{equation}

\section{Estimation of multivariate probabilistic benchmarks in noisy datasets}\label{sec:methods}

Based on Propositions 1 and 2 above, a 2-step processes is suggested to calculate multivariate probabilistic benchmarks in noisy data sets:

\begin{enumerate}
    \item In the first step, a swarm algorithm estimates the vector of parameters of a double-hyperbolic noise filter. The optimal estimates of the vector maximize the dependence structure $\theta$ in an Archimidean copula calculated with noisy KPIs. The optimal double-hyperbolic filter that maximizes $\theta$ is used to denoise the KPIs.   
    
    \item In the second step, a relevance vector machine is applied to the denoised KPIs in order to calculate multivariate probabilistic benchmarks. Besides estimating isolines of benchmarks, the relevance vector machine allows to identify factors that influence the benchmarks.
\end{enumerate}

\subsection{Step 1: Double-hyperbolic undersampling and swarm optimization}

Let $f_h  \left( \bm{\psi}, \mathbf{y} \right)$  be the real $\mathcal{R}$ part---the imaginary part is discarded---of a translated generalized hyperbola of the form \citep{hamilton1998}:
\begin{equation}
    f_h \left( \bm{\psi}, \mathbf{y} \right) :=
\mathcal{R} 
\left\{
\psi_1
\sqrt{\psi_2 + \frac{\psi_3}{\psi_4 + \mathbf{y}}}
\right\},
\quad \left\{ \psi_1, \psi_2,  \psi_3,  \psi_4  \right\} \in \bm{\psi},
\label{eq:hyperbola}
\end{equation}
If $f_h^\perp  \left( \bm{\psi ^\perp}, \mathbf{y} \right)$  is an orthogonal/quasi-orthogonal rotation of the translated generalized hyperbola defined by equation \ref{eq:hyperbola}---with rotation parameters $\left\{ \psi_1^\perp, \psi_2^\perp,  \psi_3^\perp,  \psi_4^\perp  \right\} \ni \bm{\psi}^\perp$---then the region of the double hyperbola defined by the lobes of $f_h  \left( \bm{\psi}, \mathbf{y} \right)$ and $f_h^\perp  \left( \bm{\psi ^\perp}, \mathbf{y} \right)$ can be used to filter the noise of the joint distribution of $\mathbf{y}$, if elements of  $\mathbf{y}$ outside the lobes of $f_h \left( \bm{\psi}, \mathbf{y} \right)$ and inside the lobes of the rotated hyperbola $f_h^\perp  \left( \bm{\psi ^\perp}, \mathbf{y} \right)$ are discarded.

Let $\mathbf{y}_h \subset \mathbf{y}$ be a vector with the non-discarded elements of $\mathbf{y}$ inside the lobes of $f_h \left( \bm{\psi}, \mathbf{y} \right)$ and outside the lobes of $f_h^\perp  \left( \bm{\psi ^\perp}, y \right)$. The vector $\mathbf{y}_h$ is an optimal noise reduction of the original data $\mathbf{y}$ if the values of $\bm{\psi}$ and $\bm{\psi}^\perp$  maximize the dependence structure ($\theta$) of an Archimidean copula estimated with \textit{samples} of $\mathbf{y}$, 
\begin{equation}
     \max_{
     \substack{
          \left\{
     \bm{\psi}, \bm{\psi^\perp} 
     \right\}
     \in \mathbb{R} \\  \mathbf{y}_h \subset \mathbf{y} 
     }
     }
       \left( -2 \int_0^1 \frac{u^ {-\theta} - 1 }{-\theta u^ {-(\theta + 1)}}
    \right)^{-1} - 2 \int_0^1 \frac{u^ {-\theta} - 1}{-\theta u^ {-(\theta + 1)}}
    \label{eq:maxtheta}
\end{equation}

Box 1 below shows a swarm algorithm proposed to estimate the optimal values of $\bm{\psi}$ and $\bm{\psi}^\perp$ that maximize $\theta$. The algorithm maximizes the co-dependence in the Archimidean copula by taking samples of the KPIs contained in $\mathbf{y}$. The structure of the swarm algorithm---separation, alignment, cohesion---is inspired by the BOIDS algorithm of artificial life described in \citet{reynolds1987flocks}.
\smallskip

\begin{tcolorbox}[arc=0pt,boxrule=0pt,bottomrule=1pt,toprule=1pt]
\textbf{Box 1. Pseudo-code of the swarm algorithm}
\tcblower
\begin{algorithm}[H]
 \KwData{$\left\{y_1,y_2,...,y_j\right\} \ni \mathbf{y}$}
 \KwResult{$\bm{\psi}$, $\bm{\psi}^\perp$}
 initialization\;
$\delta, M, \theta_0, p_0, , p^\perp_0$, $\zeta$, $\zeta^*$  \;
\While{$m \in \mathbb{Z}_+$}{
            $w_\delta = \delta \frac{|p|}{\norm{p} }$,  $\quad w_\delta^\perp = \delta \frac{|p^\perp|}{\norm{p^\perp} } $ \;
                \For{ $m \leftarrow 1$} 
                     {
                     random exploration of hyperbola parameters\;
                    $p_m = p_{m - 1} +  w_\delta \epsilon$, $\quad p_m^\perp = p^{m - 1}_\perp +  w_\delta^\perp\epsilon$, $\quad\epsilon \sim (0,1)$ \;
                    hyperbolic undersampling\;
                   $ y_h = f_h ( p_m, \mathbf{y}) $, $\quad y_h^\perp = f_h^\perp (p_m^\perp , \mathbf{y})$, $\quad\left\{ y_h, y_h^\perp
                   \right\} \ni \mathbf{y}_h$\;
                 copula dependence estimated with filtered $m$-samples\;
                   $ \hat\theta_m = \mathcal{C}_\theta (\mathbf{y}_h) $ \;
                    }
    $\hat\theta^* = \max \left\{ \hat\theta_i \right\}_{i= 1}^m$ (optimal dependence)\;
    $p^* = p (\hat\theta^*)$,  $\quad p^{\perp *} = p^\perp (\hat\theta^*)$ (optimal hyperbola parameters)\;
    cohesion   = $\frac{1}{2} \left(  \norm{ p_m - p_m^* } + \norm{ p_m^\perp - p_m^{\perp *} } \right)$\;
    separation = $\frac{1}{2} \left(  \norm{ p_m - \overline{p}_m^* } + \norm{ p_m^\perp - \overline{p}_m^{\perp *} } \right)$\;

      \eIf{$\hat\theta^*$ > $\hat\theta^{m-1}$}{
            $\hat\theta_m = \hat\theta^*$ \;
            $p_m = p^*$, $\quad p_m^\perp = p^{\perp *}$ \;
            alignment\;
            $\delta_m = \delta_{m-1} \left( \zeta^*  \right)$ \;
            $m = M - 1$ \;
       }{
            $\delta_m =  \delta_{m-1} \left( \zeta  \right)$  \;
      }
}
\end{algorithm}
\end{tcolorbox}

In the swarm algorithm, $\delta, M, \theta_0, p_0, p^\perp_0$, $\zeta$, $\zeta^*$ are initialization parameters. The parameter $\delta \in \mathbb{R}_+$ controls the initial dispersion of the particles, $M$ is the initial number of particles used to explore possible values of $\theta$; $\theta_0 = 0$ is the starting value of $\theta_{m}$; $p_0, p^\perp_0$ are the starting values of $p_m, p^\perp_m$;  $\zeta$, $\zeta^*$  are parameters that control the degree of exploration in the swarm algorithm. Exploitation ($\delta$) and exploration parameters ($\zeta$, $\zeta^*$) are typical of metaheuristic algorithms in general and swarm intelligence in particular---see for example \citet{tilahun2019balancing}.

The algorithm described in Box 1 explores optimal values of the hyperbola parameters $p_m, p^\perp_m$ during $m$-iterations, based on two behavioral rules: cohesion and separation. Swarm cohesion depends on the euclidean norm between $p_m, p^\perp_m$ and the optimal values $p_m^*, p^{\perp*}_m$ calculated with $\theta^*$. Swarm separation is a function of the norm between $p_m, p^\perp_m$ and the centroids $\overline{p}_m^*, \overline{p}_m^{\perp *}$. Cohesion abstains the swarm $m$ from including extreme outliers---and thus avoids a biased estimation of $\theta$---and separation guarantees that the swarm properly explores all the potential values that can maximize $\theta$ for an optimal noise filtering. Alignment is achieved by gradually reducing exploration and exploitation with $\zeta^*$ ($0 < \zeta > \zeta^* \leq 1$).

\subsection{Step 2: Relevance vector machines}

Traditional methods of supervised learning---as stochastic vector machines---produce point estimates of benchmarks as an output. Relevance vector machines, in contrast, estimate the conditional distribution of multivariate benchmarks in a fully probabilistic framework. Compared to stochastic vector machines, relevance vector machines capture uncertainty and make use of a small number of kernel functions to produce posterior probabilities of membership classification.

Let $\left\{ \mathbf{x}_i \right\}_{i=1}^k$ be a $k$-set of covariates influencing the KPIs contained in $\mathbf{y}$. The importance of each covariate is defined by a weight vector $\mathbf{w} = (w_0,\dots,w_k)$. In a linear approach, $
\mathbf{y} = \mathbf{w}^\top\mathbf{x}$. In the presence of a non-linear relationship between $\mathbf{y}$ and $\mathbf{x}$, a nonlinear maping $\mathbf{x} \to \phi(\mathbf{x})$ is a basis function for $
\mathbf{y} = \mathbf{w}^\top \phi(\mathbf{x})$. 

Given an additive noise $\epsilon_k$, the benchmark targets $\mathbf{t}$ will be,
\begin{equation}
    \mathbf{t} = \mathbf{w}^\top \phi(\mathbf{x}) + \epsilon_k,
    \label{eq:targets}
\end{equation}
where $\epsilon_k$ are independent samples from a mean-zero Gaussian noise process with variance $\sigma^2$. \citet{tipping2000sparse} and
\citet{tipping2001sparse} offer a spare Bayesian learning approach to estimate $\mathbf{w}$ in Equation \ref{eq:targets} based on the likelihood of the complete data set,
\[
p(\mathbf{t} | \mathbf{w}, \sigma^2 )  = (2\pi \sigma^2) \exp \left\{ -\frac{1}{2\sigma^2} || \mathbf{t}  - \Phi \mathbf{w} ||^2
\right\},
\]
where $\Phi$ is a $k \times (k + 1)$ design matrix $\Phi = \left[ \phi(\mathbf{x}_1),\phi(\mathbf{x}_2),\dots, \phi(\mathbf{x}_k) \right]^\top$ and $\phi(\mathbf{x}_i) = \left[1, \mathcal{K}(\mathbf{x}_i,\mathbf{x}_1),\mathcal{K}(\mathbf{x}_i,\mathbf{x}_2),\dots, \mathcal{K}(\mathbf{x}_i,\mathbf{x}_k) \right]^\top$ for a kernel function $\mathcal{K}(\cdot,\cdot)$. In a zero-mean Gaussian prior for $\mathbf{w}$,
\[
    p(\mathbf{w} | \bm{\alpha}) = \prod_{i = 0}^k \mathcal{G} ( w_i | 0,\alpha_i^{-1}),
\]
$\bm{\alpha}$ is a vector of $k+1$ hyperparameters, and the posterior distribution over the weights is: 
\begin{align*}
    p(\mathbf{w} | \mathbf{t}, \bm{\alpha}, \sigma^2) & = \frac{
    p(\mathbf{t} | \mathbf{w}, \sigma^2) p(\mathbf{w} | \bm{\alpha})
    }{p(\mathbf{t} | \bm{\alpha}, \sigma^2)}, \\
    & = (2\pi)^{-(k+1)/2} |\mathbf{\Sigma}|^{-1/2} \exp \left\{ -\frac{1}{2}  
    (\mathbf{w} - \bm{\mu})^\top
    \mathbf{\Sigma}^{-1} 
    (\mathbf{w} - \bm{\mu}) ||^2
\right\},
\end{align*}
where $\mathbf{\Sigma} = (\sigma^{-2} \mathbf{\Phi}^\top \mathbf{\Phi} + \mathbf{A})^{-1}$, $\bm{\mu} = \sigma^{-2} \mathbf{\Sigma} \mathbf{\Phi}^\top \mathbf{t}$ and
$\mathbf{A} = \text{diag}(\alpha_1,\alpha_2,\dots,\alpha_k)$. Updating methods for $\alpha_i$ are described in \citet{barber2012}. The complete specification of the hierarchical priors---based on the automatic relevance determination of \citet{mackay1996bayesian} and \citet{neal2012bayesian}---can be found in \citet{tipping2001sparse}. 

The assignment of an individual hyperparameter $\alpha_k$ to each weight $w_k$ allows to achieve sparsity in the relevance vector machine. As the posterior distribution of many of the weights is peaked around zero, non-zero weights are associated only with `relevant' vectors, i.e. with the most relevant influencing factors of the probabilistic benchmarks estimated with the denoised KPIs.


\section{Empirical application: probabilistic benchmarks in nanofinance+}\label{sec:empap}

This section illustrates the methods described in Section \ref{sec:methods} with an application to a database of 7830 nanofinance+ groups receiving entepreneurship and business training in 14 African countries: Benin, Burkina Faso, Ethiopia, Ghana, Malawi, Mozambique, Niger, Sierra Leone, South Africa, Sri Lanka, Tanzania, Togo, Uganda and Zambia. Almost all of the groups in the database work with a development agency (94\%), and 43\% of the groups are located in rural regions.

Table \ref{tab:desmicmac} shows descriptive statistics of group-level characteristics and the macro-economic environment of the countries where the groups operate. On average, each member of NF+ contributes around 29 USD of savings to the common fund and receives on average a loan of 22 USD. Despite the low values of savings and loans, returns on savings in the groups are on average 47\%, whereas the equity per member is on average equal to 40 USD (Table \ref{tab:desmicmac}). 

Returns on savings ($y_1$) and equity per member ($y_2$) are the KPIs used for calculating the benchmarks of NF+ in the empirical application. Hence, $j =2$, $\mathbf{y} = [y_1 \;\; y_2]$, and the joint distribution in Proposition 1 simplifies to,
\begin{equation}
    f_{y_1,y_2} \left( y_1, y_2\right) = f_{y_1|y_2} \left( y_1|y_2\right) f_{y_1|y_2} \left( y_2\right) = f_{y_2|y_1} \left( y_2|y_1\right) f_{y_2|y_1} \left( y_1\right).
    \label{eq:joint2}
\end{equation}
Successful units---NF+ groups with a higher financial performance---will be those with KPIs delimitied by the isolines of the threshold $\tau$,
\[
1 - \int_{\tau_1}^{\infty} \int_{\tau_2}^{\infty} f_{y_1,y_2} \left( y_1, y_2\right) d y_1 d y_2,
\]
for a probabilistic benchmark $\tau \in \left\{\tau_1 \;\; \tau_2\right\}$, $\tau \in \mathbb{R}^2$.

Following Proposition 2, the joint density of the KPIs (equation \ref{eq:joint2}) is approximated with a bivariate Archimedean copula:
\begin{equation}
   \mathcal{C}_g (u_1, u_2) =
    \begin{cases}
        g^{-1}(g(u_1) + g(u_2)) \text{ if } g(u_1) + g(u_2) \leq g(0) \\
        0 \text{ otherwise. }
    \end{cases}
\end{equation}
Clayton's Archimedean copula is particularly suitable to model the dynamics of nanofinance+. Clayton's copula has greater dependence in the lower tail compared to the upper tail. In the case of NF+, greater lower tail dependence is expected because groups with low equity will have zero or negative returns, while in contrast there is more dispersion in the indicators of groups with higher performance---i.e. some groups show higher equity but low levels of returns due to lower repayment rates, while groups with low equity may have higher returns due to the higher interest rates charged for their loans.

A bivariate Clayton's Archimedean copula for the uniform marginal distributions of returns on savings ($u_1$) and equity per member ($u_2$) will be:
\begin{align}
    \mathcal{C}_{\theta} (u_1, u_2) & = g^{-1}(g(u_1) + g(u_2)) \\
     & = (1 + u_1^{-\theta} - 1 +  u_2^{-\theta} -1)^{-1/\theta} \\
      & = (u_1^{-\theta} +  u_2^{-\theta} -1)^{-1/\theta}
\end{align}
with a probability density function,
\begin{equation}
    c_{\theta} (u_1, u_2)  = \frac{\partial^2}{\partial u_1 \partial u_2} \mathcal{C}_{\theta} = \frac{1 + \theta}{(u_1 u_2) ^{\theta + 1} } (u_1^{-\theta} +  u_2^{-\theta} -1)^{-2 - \frac{1}{\theta}}
\end{equation}
and a co-dependence  parameter $\theta \in [0,+\infty)$,
\begin{equation}
    \theta = \left( -2 \int_0^1 \frac{u^ {-\theta} - 1 }{-\theta u^ {-(\theta + 1)}}
    \right)^{-1} - 2 \int_0^1 \frac{u^ {-\theta} - 1}{-\theta u^ {-(\theta + 1)}}.
\end{equation}
The parameter $\theta$ controls the amount of dependence in $\mathcal{C}_{\theta} (u_1, u_2)$. When $\theta \to +\infty$ the dependency between $u_1$ and $u_2$ approaches comonoticity,
\begin{equation}
    \lim_{\theta \to +\infty} \mathcal{C}_{\theta} (u_1, u_2) = \min (u_1, u_2),
\end{equation}
while in turn when $\theta \to 0$, $u_1$ and $u_2$ become independent:
\begin{equation}
    \lim_{\theta \to 0} \mathcal{C}_{\theta} (u_1, u_2) = u_1 u_2.
\end{equation}
In the case of returns on savings and equity per member, it is expected that $\theta \to +\infty$, as both financial indicators should show lower tail co-dependence in NF+.

\begin{figure}[ht]
\centering
\caption{Swarm optimisation of $\theta$ in Clay's Archimidean copula. The copula was estimated with the data of returns on savings and equity per member of nanofinance groups. In the graph, the swarm shows greater dispersion at the start of the iterations, but the cohesion and separation of the flock converge after the iteration 15, when the value of the estimate of $\theta$ tends to stabilize.}
\includegraphics[width=\textwidth]{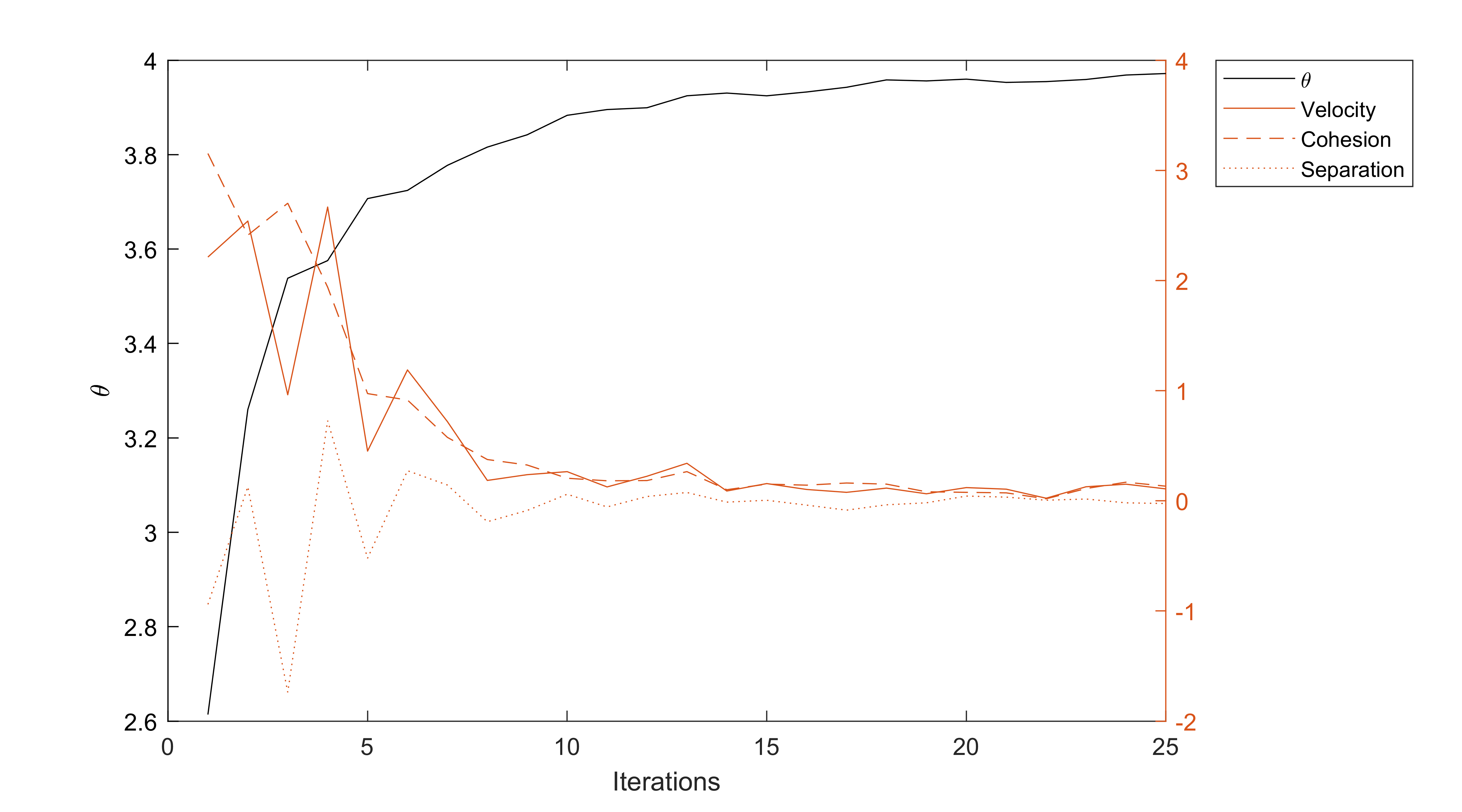}
\label{fig:swarm}
\end{figure}

\begin{figure}[ht]
\centering
\caption{Denoising with double-hyperbolic undersampling. For an optimal filtering of noise, the points outside the lobes of the first hyperbole are discarded in graph (b), and the points inside the lobes of the second hyperbole are discarded in graph (d). Figure (e) shows the relationship between the KPIs before denoising, and figure (f) shows the relation after denoising.}
\includegraphics[width=\textwidth]{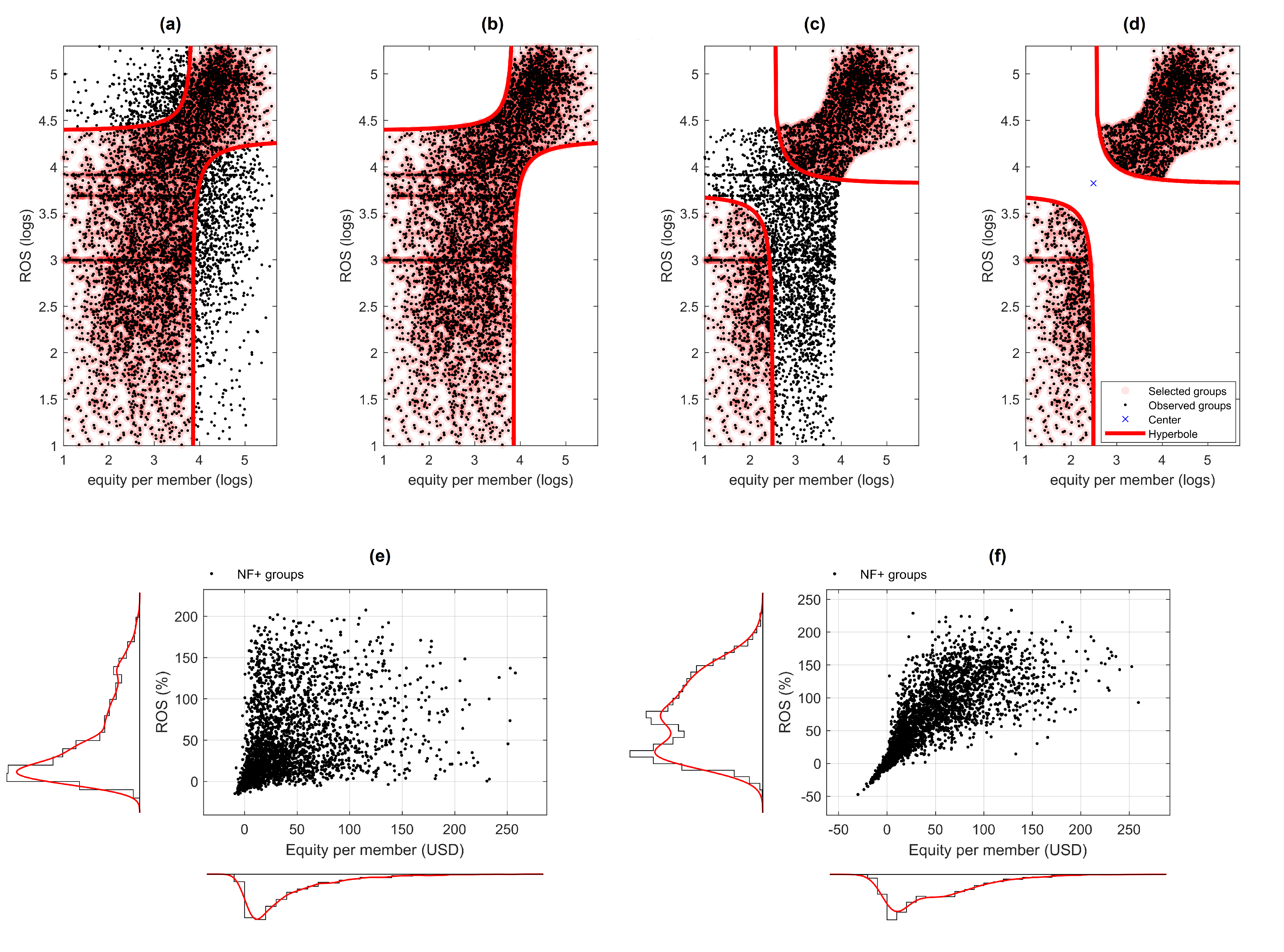}
\label{fig:filtering}
\end{figure}

Figure \ref{fig:swarm} shows indeed that the swarm optimization of $\theta$---using the data of returns on savings and equity per member---leads to a value of $\hat\theta = 3.97$. The estimates of the parameters of the hyperbolas for $\hat\theta$ are equal to,
\begin{align*}
    \hat{\bm\psi} & := \left\{ \hat{\psi}_1, \hat{\psi}_2, \hat{\psi}_3, \hat{\psi}_4 \right\} = \left\{ 77.42, 0.87, -10.38, -46.51 \right\}, \\
    \hat{\bm\psi}^\perp & := \left\{ \hat{\psi}_1^\perp, \hat{\psi}_2^\perp, \hat{\psi}_3^\perp, \hat{\psi}_4^\perp \right\} = \left\{ 55.92, 0.67, 2.26, -15.43 \right\}.
\end{align*}
Figure \ref{fig:filtering} shows the optimal denoising of the KPIs of NF+ with double-hyperbolic undersampling. The first step discards the values of ROS and EPM outside the lobes of the hyperbole estimated with $\bm\psi$ and inside the lobes of the hyperbole estimated with $\bm{\psi^\perp}$ (Figures \ref{fig:filtering}b and \ref{fig:filtering}d). The co-dependence between the KPIs before denoising is contaminated with a high number of outliers (Figure \ref{fig:filtering}e). After denoising, the co-dependence in the lower and upper tails of the KPIs is kept but noisy elements are discarded (Figure \ref{fig:filtering}f).

\begin{figure}[ht]
\centering
\caption{Probabilistic benchmarks estimated with the relevance vector machine}
\includegraphics[width=\textwidth]{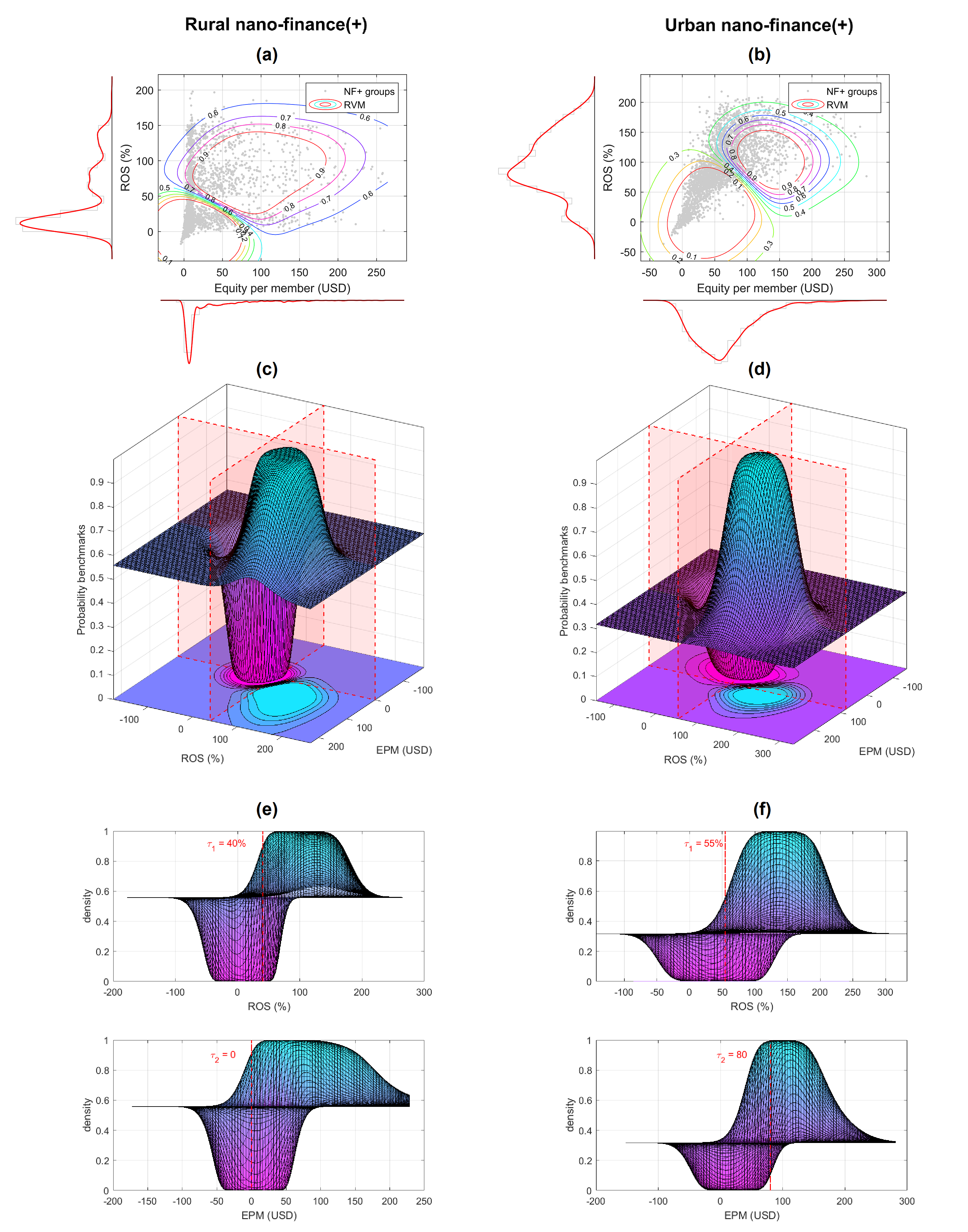}
\label{fig:RVM_surfs}
\end{figure}

Table \ref{tab:results} and Figure \ref{fig:RVM_surfs} show the results of estimating the relevance vector machine with the denoised KPIs (step 2). In terms of continuous factors influencing the benchmarks, the main covariates affecting the financial benchmarks of NF+ are those related to the macroeconomic environment, mainly GDP growth, poverty, inequality and the percentage of rural population in the country where a NF+ group operates (Table \ref{tab:results}). Savings accumulation and loan provision are the main group-level characteristics influencing the financial benchmarks of NF+; this result is expected---because in NF+ the lending channel is the main source of profit generation---and shows the ability of the relevance vector machine to properly detect variables related to financial benchmarks in denoised datasets. 

In relation to categorical factors influencing the benchmarks, Figure \ref{fig:RVM_surfs} shows that the probabilistic benchmarks of NF+ are different in rural groups (Figure \ref{fig:RVM_surfs} left) compared to urban groups (Figure \ref{fig:RVM_surfs} right). While both rural and urban groups have a concentration of financial performance in the lower tail of the joint distribution of the KPIs, higher dispersion in the upper tail is observed in rural groups, and hence the isolines of the probabilistic benchmarks are wider for rural groups compared to urban groups. 

In the case of urban and peri-urban nano-finance, groups can be classified as successful with a probability higher than 90\% (red contour isoline in Figure \ref{fig:RVM_surfs}b) when the groups have returns higher than 55\% and equity higher than 80 USD per member (Figures \ref{fig:RVM_surfs}f). In rural NF+, however, groups that do not show negative returns and have an equity per member higher than 10 USD are classified as successful with a probability higher than 80\% (Figures  \ref{fig:RVM_surfs}c and \ref{fig:RVM_surfs}e).

\renewcommand{\arraystretch}{1.15}

\begin{table}[htbp]
  \centering
  \caption{Descriptive statistics of the SAVIX. Nanofinance groups in the SAVIX have on average 21 members and 82\% of the members are women. The members show a high commitment to the group meetings: member's attendance is 92\%, and the members that end up leaving the group are only 1.2\% of the total of participants. In macro-economic terms, the GDP growth in the countries where the nanofinance groups operate is on average 4.88\%, and the GDP per capita is on average 1353 USD. The countries where the groups are located have also low levels of literacy (the literacy rate is 56\%), low levels of financial inclusion (the indicator of financial deepening is 33\%), and a high percentage of population living in poverty (40\%) and in rural areas (60\%).}
    \begin{tabular}{lrrrr}
    \textbf{Variables} & \multicolumn{1}{c}{\textbf{Mean}} & \multicolumn{1}{c}{\textbf{Std. Dev.}} & \multicolumn{1}{c}{\textbf{Min}} & \multicolumn{1}{c}{\textbf{Max}} \\
    \midrule
    \multicolumn{5}{l}{\textbf{Group-level characteristics of nanofinance+}} \\   [.05in]
    Returns on savings$^a$ & 48.63 & 47.14 & 0  & 199.47 \\
    Equity per member$^b$ & 40.41 & 40.25 & 0.10  & 269.90 \\
    Savings per member$^b$ & 29.15 & 28.71 & 0.06  & 235.79 \\
    Fund utilisation rate$^b$ & 57.73 & 34.88 & 0  & 100.00 \\
    Number of loans per member & 0.51  & 0.33  & 0  & 1.00 \\
    Average loans per member$^b$ & 22.40 & 29.51 & 0  & 186.14 \\
    Welfare fund per member$^b$ & 1.32  & 1.67  & 0 & 12.59 \\
    Member's attendance$^a$ & 92.32 & 11.16 & 39.29 & 100.00 \\
    Drop-out rate$^a$ & 1.17  & 4.32  & 0  & 45.00 \\
    Number of members & 21.11  & 6.55  & 5  & 33.50 \\
    Women members$^a$ & 81.99 & 23.19 & 0  & 100.00 \\
    Accumulated loans per member & 0.51  & 0.33  & 0.00  & 1.75 \\ \midrule 
\multicolumn{5}{l}{\textbf{Macro-economic variables}}\\  [.05in]
    Uncertainty (inflation deviation)$^a$ & 2.87  & 1.39  & 0.66  & 11.54 \\
    Inflation rate$^a$ & 6.68  & 6.88  & -1.01 & 21.87 \\
    Age-dependency ratio$^a$ & 87.19 & 12.91 & 51.23 & 111.67 \\
    Gini coefficient$^a$  & 45.40 & 8.10  & 32.90 & 63.20 \\
    Financial deepening$^a$ & 33.31 & 31.75 & 12.55 & 179.78 \\
    Literacy rate$^a$ & 56.24 & 24.18 & 15.46 & 94.37 \\
    GDP per capita$^b$ & 1353.04 & 1410.32 & 386.73 & 7575.18 \\
    Population density$^a$ & 82.61 & 54.79 & 15.12 & 334.33 \\
    Rural population$^a$ & 60.16 & 13.23 & 34.15 & 84.03 \\
    Poverty headcount ratio$^a$ & 39.18 & 11.54 & 17.70 & 56.90 \\
    GDP growth$^a$ & 4.88  & 1.61  & -1.93 & 10.25 \\
    \bottomrule
    \multicolumn{5}{l}{\scriptsize{$^a$ Percentage (\%)}}\\[-.05in]
    \multicolumn{5}{l}{\scriptsize{$^b$ US dollars (USD)}}
    \end{tabular}%
  \label{tab:desmicmac}%
\end{table}%

\begin{table}[htbp]
\small
  \centering
  \caption{Results of estimating the relevance vector machine for $\mathbf{y}$ with the set of covariates $\mathbf{x}$}
    \begin{tabular}{crrrrrr}
    \textbf{Type} & \multicolumn{1}{c}{\textbf{Covariates ($\mathbf{x}$)}} & \multicolumn{1}{c}{\textbf{AUC$^a$}} & \multicolumn{1}{c}{\textbf{Gini}} & \multicolumn{1}{c}{\textbf{Bacc$^b$}} & \multicolumn{1}{c}{\textbf{Prec$^c$}} & \multicolumn{1}{c}{\textbf{FDR$^d$}} \\

    \midrule
    \multicolumn{1}{c}{\multirow{9}[2]{*}{
    \shortstack[c]{
    Micro-level \\ characteristics
    }
    }} & Savings per member* & 1.0000 & 1.0000 & 0.9908 & 1.0000 & 0.0000 \\
      & Fund utilization rate & 0.7096 & 0.4193 & 0.6326 & 0.3723 & 0.6277 \\
      & Number of loans per member* & 0.8261 & 0.6522 & 0.7514 & 0.6355 & 0.3645 \\
      & Average loans per member* & 0.7852 & 0.5703 & 0.8716 & 0.9955 & 0.0045 \\
      & Welfare fund per member* & 0.8035 & 0.6071 & 0.7675 & 0.8110 & 0.1890 \\
      & Mmember's attendance & 0.6067 & 0.2134 & 0.5000 & 0.0000 & 1.0000 \\
      & Drop-out rate & 0.5511 & 0.1021 & 0.5149 & 0.0731 & 0.9269 \\
      & Women members & 0.6374 & 0.2748 & 0.5155 & 0.0619 & 0.9381 \\
      & Accumulated loans per member* & 0.8261 & 0.6522 & 0.7514 & 0.6355 & 0.3645 \\
      & Rural location* & 0.7946 & 0.5893 & 0.7946 & 0.8031 & 0.1969 \\
    \midrule
    \multicolumn{1}{c}{\multirow{11}[1]{*}{
    \shortstack[c]{
      Macro-economic \\ variables
    }
    }} & Uncertainty (inflation deviation)* & 0.7620 & 0.5241 & 0.7128 & 0.5073 & 0.4927 \\
      & Inflation rate & 0.5860 & 0.1721 & 0.5000 & 0.0000 & 1.0000 \\
      & Age-dependency ratio* & 0.7606 & 0.5212 & 0.7836 & 0.8268 & 0.1732 \\
      & Inequality (Gini index)* & 0.8393 & 0.6785 & 0.7447 & 0.5534 & 0.4466 \\
      & Financial deepening & 0.7369 & 0.4739 & 0.7378 & 0.5816 & 0.4184 \\
      & Literacy rate & 0.6774 & 0.3548 & 0.5000 & 0.0000 & 1.0000 \\
      & GDP per capita* & 0.7873 & 0.5745 & 0.5011 & 0.1271 & 0.8729 \\
      & Population density* & 0.7939 & 0.5878 & 0.7276 & 0.5748 & 0.4252 \\
      & Rural population in a country* & 0.8485 & 0.6970 & 0.7532 & 0.5591 & 0.4409 \\
      & Poverty headcount ratio* & 0.8487 & 0.6973 & 0.7961 & 0.7030 & 0.2970 \\
      & GDP growth* & 0.8516 & 0.7031 & 0.7374 & 0.6614 & 0.3386 \\
    \midrule
    \multicolumn{1}{c}{\multirow{8}[1]{*}{ 
    \shortstack[c]{
   Facilitation \\ mechanisms \\ of development \\ agencies
    }
    }} & No facilitating agency & 0.5025 & 0.0051 & 0.5000 & 0.0000 & 1.0000 \\
      & No donors & 0.5479 & 0.0957 & 0.5000 & 0.0000 & 1.0000 \\
      & Group formed by paid agent & 0.6803 & 0.3605 & 0.6803 & 0.6828 & 0.3172 \\
      & Group formed by field officer & 0.5147 & 0.0295 & 0.5000 & 0.0000 & 1.0000 \\
      & Group formed by unpaid agent & 0.5264 & 0.0529 & 0.5264 & 0.0754 & 0.9246 \\
      & Group formed by project-paid agent & 0.5264 & 0.0528 & 0.5264 & 0.0877 & 0.9123 \\
      & Graduated groups & 0.5882 & 0.1764 & 0.5000 & 0.0000 & 1.0000 \\
\bottomrule

    \multicolumn{7}{l}{\scriptsize{(*) Variables with the best machine-learning indicators}} \\[-.05in]
    \multicolumn{7}{l}{\scriptsize{$^a$ AUC: Area under the ROC courve}} \\[-.05in]
    \multicolumn{7}{l}{\scriptsize{$^b$ Bacc: Balanced accuracy}} \\[-.05in]
    \multicolumn{7}{l}{\scriptsize{$^c$ Prec: Precision}} \\[-.05in]
    \multicolumn{7}{l}{\scriptsize{$^d$ FDR: False detection rate}} \\
    \end{tabular}%
  \label{tab:results}%
\end{table}%

\FloatBarrier

\section{Conclusion}\label{sec:conclusion}

This study suggested a 2-step approach for calculating probabilistic benchmarks with noisy KPIs. An empirical application to a noisy database of nanofinance+ shows that the methods are able to denoise KPIs, estimate probabilistic benchmarks, and properly identify the continuous and discrete factors influencing the benchmarks. 

In the case of NF+ groups with business training, the results indicate that macroeconomic factors and the region where a group is located influence their financial benchmarks. Governments, international donors and development agencies can use the estimated benchmarks for monitoring the performance of NF+ and gain an independent perspective about how well a group/project is performing when compared to other similar groups/projects. In the presence of performance gaps, the benchmarks will be useful to identify opportunities for change and improvement among the groups\footnote{It is estimated that over 100 million people in 10.5 million households participate in nanofinance groups worldwide (\cite{greaney2016}; \cite{burlando-2017}).
Due to the importance of NF+ for financial inclusion and multidimensional poverty reduction, all major international donors and development agencies work with NF+, but these organizations lack of benchmarks to evaluate the financial performance of NF+ groups.}.

Future studies can extend the denoising methods to the quadratic surface defined by hyperbolic cylinders. The higher-dimensional hierarchical Archimedean copula proposed by \citet{savu2010} can be applied to approximate the multivariate probability distribution of KPIs denoised with hyperbolic cylinders. The recent developments in orthogonal machine learning---see \textit{inter alia} \citet{oprescu2018orthogonal}, \citet{knaus2018double}, \citet{semenova2018essays} or \citet{kreif2019machine}---can be used to estimate quasi-causal factors influencing the benchmarsk, complementing the non-parametric correlational approach of relevance vector machines.

\printbibliography


\end{document}